\documentclass[a4paper,12pt]{article} 
\pdfoutput=1
\usepackage{jcappub}

\usepackage[utf8]{inputenc}
\usepackage{dsfont} 
\usepackage[english]{babel}
\usepackage{bm}
\usepackage{dsfont} 
\usepackage{bbm}
\usepackage{subfigure}
\usepackage{bbm}
\usepackage{ulem}

\newcommand{\eref}[1]{Eq.~(\ref{#1})}
\newcommand{\fref}[1]{Figure~\ref{#1}}
\newcommand{\sref}[1]{Section~\ref{#1}}

\title{
Stability of three neutrino flavor conversion in supernovae
}

\author[a]{{\large Christian D\"oring,}}
\author[a,b]{{\large Rasmus S. L. Hansen}}
\author[a]{{\large and Manfred Lindner}}

\affiliation[a]{Max-Planck-Institut f\"ur Kernphysik,\\ Saupfercheckweg 1, 69117 Heidelberg, Germany}
\affiliation[b]{Department of Physics and Astronomy, Aarhus University, \\ 
Ny Munkegade 120, 8000 Aarhus C, Denmark}
\emailAdd{cdoering@mpi-hd.mpg.de}
\emailAdd{rshansen@phys.au.dk}
\emailAdd{lindner@mpi-hd.mpg.de}

\abstract{
Neutrino-neutrino interactions can lead to collective flavor conversion in the dense parts of a core collapse supernova. Growing instabilities that lead to collective conversions have been studied intensely in the limit of two-neutrino species and occur for inverted mass ordering in the case of a perfectly spherical supernova. We examine two simple models of colliding and intersecting neutrino beams and show, that for three neutrino species instabilities exist also for normal mass ordering even in the case of a fully symmetric system. Whereas the instability for inverted mass ordering is associated with $\Delta m_{31}^2$, the new instability we find for normal mass ordering is associated with $\Delta m_{21}^2$. As a consequence,  the growth rate of these new instabilities for normal ordering is smaller by about an order of magnitude compared to the rates of the well studied case of inverted ordering.
}

\begin{document}

\notoc
\maketitle

\section{Introduction}
\label{sec:intro}
Core collapse supernovae (SN) are powerful astrophysical events that involve the interplay of various physical fields. Next to light and gravitational waves signals, a galactic SN would also produce a prominent neutrino signal leading to the possibility of studying the influence of fundamental physics on astrophysical-scale events.
Neutrinos are mainly produced by charged current $\beta$-processes and pair production during the first ten seconds after the onset of the collapse and have energies of order tens of MeV. They carry away $99\%$  of the gravitational binding energy released in a SN and are expected to play a major role in the dynamics of the SN explosion by revitalising the stalled shock wave and driving nucleosynthesis of heavy elements.

Both reviving the shock wave and the nucleosynthesis depend on the flavor composition of the neutrinos which is affected by vacuum oscillation and the Mikheyev-Smirnov-Wolfenstein (MSW) matter effect~\cite{Wolfenstein:1977ue,Mikheev:1986gs,Mikheev:1986wj}. Due to the high neutrino densities of $10^{30-35}$ per $\textrm{cm}^3$ close to the neutrino sphere (radius $\sim 50\,\textrm{km}$) during the accretion phase, neutrino-neutrino interactions must also be taken into account. However, this makes the neutrino flavour evolution non-linear and results in what is called collective oscillations, where neutrinos of different energies all convert from one flavour to another at the same time. Since collective oscillations significantly affect the flavor composition, they can also have an important impact on the shock wave revival as well as on nucleosynthesis.

The physics of dense neutrino media are also relevant in the environment of neutron star mergers and in early Universe cosmology. The importance of neutrino-neutrino interactions was first realised by Pantaleone \cite{Pantaleone:1992eq}. The corresponding quantum kinetic equations were derived by Sigl and Raffelt \cite{Sigl:1992fn}, and these non-linear equations have since been intensively studied and shown to give rise to collective oscillations such as synchronisation \cite{Pastor:2001iu} and various forms of bipolar oscillations \cite{Duan:2005cp,Duan:2006an,Hannestad:2006nj}. Also fast flavor conversions have been found for which the typical conversion scale is proportional to the neutrino density and is not directly related to the scale of the neutrino mass difference \cite{Sawyer:2005jk,Sawyer:2008zs,Sawyer:2015dsa,Chakraborty:2016lct,Dasgupta:2016dbv,Capozzi:2017gqd,Izaguirre:2016gsx,Abbar:2017pkh}. For recent reviews of collective oscillations and supernova neutrino physics, see e.g.~\cite{Mirizzi:2015eza,Chakraborty:2016yeg}.

A powerful method to identify models where collective oscillations are present is to use a linearised stability analysis~\cite{Banerjee:2011fj}. If bipolar oscillations or fast flavour conversions are present, they are seen as unstable modes that grow exponentially. With this method, several different types of collective instabilities have been identified in two-neutrino models. For a homogeneous and isotropic neutrino gas, the simple bipolar oscillations are found for inverted mass ordering (IO), while no instability is found for normal mass ordering (NO)~\cite{Duan:2005cp,Duan:2006an,Hannestad:2006nj}. When inhomogeneities and anisotropy are allowed, other instabilities are identified such as the multi-zenith angle (MZA) instability, the multi-azimuthal angle (MAA) instability~\cite{Raffelt:2013rqa,Raffelt:2013isa} and also the fast flavour conversion~\cite{Sawyer:2005jk,Sawyer:2008zs,Sawyer:2015dsa,Chakraborty:2016lct,Dasgupta:2016dbv,Capozzi:2017gqd,Izaguirre:2016gsx,Abbar:2017pkh}. The MZA and MAA are only found for NO, while fast flavour conversions are largely independent on mixing parameters and are present for both mass orderings.
Three-neutrino effects have been included in a number of numerical studies~\cite{Duan:2007sh,Balantekin:2007es,EstebanPretel:2007yq,Gava:2008rp, Fogli:2008fj, Friedland:2010sc,Cherry:2010yc,Mirizzi:2010uz}. Although most analytical work has been done in the limit of two flavors in which muon and tau neutrinos are combined to $\nu_x$, three flavor effects have also been considered in several papers~\cite{Duan:2008za,Dasgupta:2007ws, Kneller:2009vd, Dasgupta:2010ae, Airen:2018nvp}. 
Most recently, Airen et al.~\cite{Airen:2018nvp} did an extensive normal-mode analysis including a discussion of three-neutrino effects. Noting that the neutrino is almost in a flavour eigenstate when the electron background density is large, they neglected the off-diagonal part of the vacuum-contribution to the Hamiltonian describing the flavour evolution as it is usually done (see e.g.~\cite{Duan:2006an,Duan:2005cp,Hannestad:2006nj, Banerjee:2011fj}). This allowed them to simplify the problem to two-neutrino oscillations and collect the solutions previously described in the literature in one unified framework. In the description of three-neutrino effects, they notice terms coupling the different off-diagonal parts of the density matrix through the vacuum term, but do not pursue them further. 

In our study, we include the full expression of the vacuum term in a two- and four-beam model, and numerically identify the flavour instabilities including one that has evaded attention until now. 
In \sref{sec:eom} we introduce the formalism that is used to describe flavor conversion in dense neutrino environments. Then, in \sref{sec:asol}, we construct an approximated solution to the problem. We apply this solution in \sref{sec:models} to a system of a neutrino beam colliding with an antineutrino beam and extend this model to a system of four neutrino and antineutrino beams with an nonvanishing intersection angle. The numerical results are interpreted in terms of two-neutrino instabilities in \sref{sec:PropStates} before we conclude in \sref{sec:conclusion}.

\section{Equation of Motion}
\label{sec:eom}
\subsection{Basic description}
Our goal is to describe the evolution of the flavor composition of a neutrino ensemble in a supernova. For a homogeneous ensemble of collisonless neutrinos with momentum $\textbf{p}$ and flavor $\alpha\in\{e,\mu,\tau\}$, the flavor composition is characterized by the expectation value of bilinear combinations of neutrino and antineutrino annihilation $a_{\alpha}^{\dagger}(\textbf{p}),b_{\alpha}^{\dagger}(\textbf{p})$ and creation operators $a_{\alpha}(\textbf{p}),b_{\alpha}(\textbf{p})$ \cite{Notzold:1987ik,Sigl:1992fn}. This in turn translates into a mean field density matrix formulation of the system due to the relations
\begin{gather}
 \langle a_{\beta}^{\dagger}(t,\textbf{p})a_{\alpha}(t,\textbf{p}^{\prime})\rangle=(2\pi)^3\,\delta^{(3)}(\textbf{p}-\textbf{p}^{\prime})(\rho(t,\textbf{p}))_{\alpha\beta},\\[0.3cm]
  \langle b_{\alpha}^{\dagger}(t,\textbf{p})b_{\beta}(t,\textbf{p}^{\prime})\rangle=(2\pi)^3\,\delta^{(3)}(\textbf{p}-\textbf{p}^{\prime})(\overline{\rho}(t,\textbf{p}))_{\alpha\beta}.
\end{gather}
The time evolution of the flavor composition of a neutrino ensemble can thus be described by the evolution of the mean field density matrix in three dimensional flavor space
\begin{align}
 \rho(t,\textbf{p}):=\begin{pmatrix}
                                   \rho_{ee}(t,\textbf{p}) & \rho_{e\mu}(t,\textbf{p}) & \rho_{e\tau}(t,\textbf{p})\\
                                   \rho_{\mu e}(t,\textbf{p}) & \rho_{\mu\mu}(t,\textbf{p}) & \rho_{\mu\tau}(t,\textbf{p})\\
                                   \rho_{\tau e}(t,\textbf{p}) &\rho_{\tau\mu}(t,\textbf{p})&\rho_{\tau\tau}(t,\textbf{p})
                                  \end{pmatrix},
\end{align}
where the diagonal entries encode the flavor occupation numbers and the off-diagonal elements represent flavor correlations. Furthermore we treat neutrinos as ultra-relativistic which implies that their four velocity is $v^{\mu}=(1,\textbf{v})$ and hence  $|\textbf{v}|=1$. For the sake of a more compact notation, we denote antineutrino states by negative phase-space densities  $-\overline{\rho}(E)=\rho(-E)$ with $E=|\textbf{p}|$.  The direction of motion, the energy, and the $\nu/\overline{\nu}$-nature are summarized in a mode index $i$. In a physical situation many different modes may exist and it is the interplay among those modes that is the root of collective oscillation effects.
Our interest is to investigate the time-dependent behavior of the flavor composition of the neutrino ensemble and hence of the density matrix. The density matrix evolves in flavor space for each mode according to Von Neumann's equation
\begin{gather}
 i\,\partial_t\rho_i(t,\textbf{v})=[H_i(t,\textbf{v}),\rho_i(t,\textbf{v})],
 \label{eq:VonNeumann}
\end{gather}
when collisions are neglected and applies to pure as well as to mixed states \cite{Stirner:2018ojk}. 
The Hamiltonian $H_i(t,\textbf{v})$ on the right hand side decomposes into three components that trigger neutrino oscillation
\begin{gather}
 H_i(t,\textbf{v})=H_{i}^{\text{vac}}+H^{\lambda}+H_i^{\nu\nu}(\textbf{v}).
 \label{eq:Hamiltonian}
\end{gather}
Apart from terms proportional to the unit matrix, the vacuum term for a neutrino mode of energy $E_i$ has in flavor basis the form
\begin{gather}
 H_{i}^{\text{vac}}:=\frac{U M^2U^{\dagger}}{2E_i},
\end{gather}
where $M^2:=\text{diag}(0,\Delta m_{21}^2,\Delta m_{31}^2)$ is the matrix of neutrino square mass differences and $U=R_{23}R_{13}R_{12}$ is, up to a CP-violating phase, the neutrino mixing matrix with the rotation matrices $R_{kl}$ between the $k$th and $l$th mass eigenstate $\Delta m_{kl}:=m_k^2-m_l^2,\,\, k,l\in\{1,2,3\}$ with $m_k$ being the mass of neutrino mass eigenstate $\nu_k$. The rotation matrices are
\begin{align}
R_{12}:=\begin{pmatrix}
            c_{12} &s_{12} &0\\
            -s_{12} &c_{12}&0\\
            0& 0& 1
          \end{pmatrix},\,
R_{13}:=\begin{pmatrix}
           c_{13} & 0 & s_{13}\\
           0 & 1 & 0\\
           -s_{13}&0&c_{13}
          \end{pmatrix},\,
R_{23}:=\begin{pmatrix}
          1 & 0 & 0\\
          0 &c_{23} &s_{23}\\
          0 & -s_{23} &c_{23}
          \end{pmatrix}.
\end{align}
We use here and in the following the shorthand $c_{kl}:=\cos(\theta_{kl})$ with the mixing angles $\theta_{12}, \theta_{23}$ and $\theta_{13}$. 
Note, that a positive sign of $\Delta m_{31}^2$  refers to NO while a negative sign refers to IO.

The second term in \eref{eq:Hamiltonian} describes matter effects which includes the MSW-effect and is proportional to the Fermi coupling constant $G_F$. In the weak eigenstate basis, the interaction of neutrinos with charged leptons of net density $N_{\alpha}$, $\alpha\in\{e,\mu,\tau\}$ reads
\begin{gather}
 H^{\lambda}:=\sqrt{2}G_F\,\text{diag}(N_e,N_{\mu},N_{\tau}).
\end{gather}

The interaction among neutrino modes requires high neutrino densities and also affects the flavor evolution of the density matrix. The description of  interacting neutrino modes is encoded in the last term of \eref{eq:Hamiltonian} which for a discrete set of $N$ modes with velocities $\textbf{v}_j$, $j\in\{1,...,N\}$, is given as
\begin{gather}
 H_i^{\nu\nu} := \sqrt{2}G_F n_{\nu}\sum_{j=1}^N(1-\textbf{v}_i\textbf{v}_j)\,\rho_j.
\end{gather}
Note that the Hamiltonian depends on the relative angle of the propagation directions of the modes and is proportional to the effective neutrino density $n_{\nu}:=\frac{1}{2}(n_{\nu_e}-n_{\bar{\nu}_e}+n_{\nu_x}-n_{\bar{\nu}_x}+n_{\nu_y}-n_{\bar{\nu}_y})$. 

\subsection{Two-Neutrino Limit}
In the context of supernovae, it is expected that $\nu_{\mu}$ and $\nu_{\tau}$ have almost identical number density spectra which allows for a basis transformation in three dimensional flavor space spanned by $\{|\nu_e\rangle,|\nu_{\mu}\rangle,|\nu_{\tau}\rangle\}$ such that $\nu_{\mu}$ and $\nu_{\tau}$ unmix \cite{Dasgupta:2007ws,Dasgupta:2010ae} 
\begin{gather}
 \begin{pmatrix}
  \nu_e\\
  \nu_x\\
  \nu_y
 \end{pmatrix}:=
 R_{23}^{\dagger}
 \begin{pmatrix}
  \nu_e\\
  \nu_{\mu}\\
  \nu_{\tau}
 \end{pmatrix}.
\end{gather}
In most works, collective neutrino oscillations are studied in the two-neutrino subspace spanned by $\{|\nu_e\rangle,|\nu_x\rangle\}$. This requires that $|\nu_y\rangle$ is proportional to the third mass eigenstate $|m_3\rangle$ only which can be achieved exactly for $\theta_{13}=0$ \cite{Balantekin:1999dx}. Since $\theta_{13}\approx 8^{\circ}$, the two-neutrino limit is only approximatively achievable.

In this work, we will assume that $\nu_{\mu}$ and $\nu_{\tau}$  have identical distributions and hence work in the new basis spanned by $|\nu_e\rangle, |\nu_x\rangle$ and $|\nu_y\rangle$ in order to simplify the equations. For \eref{eq:VonNeumann} this statement is equivalent to set $\theta_{23}=0$.

\section{Linearized equations}
\label{sec:asol}

\subsection{Three neutrino formalism}
Since the Von Neumann equation conserves the trace of the density matrix, we subtract the trace and split the entries into the neutrino density of the $i$th mode $g_i$ and parametrize the remaining fractions  as
\begin{align}
 \rho_i(t,\textbf{v})-\frac{1}{3}\text{Tr}\rho_i\mathbbm{1}=:\frac{g_i}{3}\begin{pmatrix} \mathfrak{s}_{1}^i(t,\textbf{v}) & \mathcal{S}_{ex}^i(t,\textbf{v}) & \mathcal{S}_{ey}^i(t,\textbf{v})\\ \mathcal{S}_{ex}^{\ast i}(t,\textbf{v}) &  \mathfrak{s}_{2}^i(t,\textbf{v}) & \mathcal{S}_{xy}^i(t,\textbf{v}) \\ \mathcal{S}_{ey}^{\ast i}(t,\textbf{v}) & \mathcal{S}_{xy}^{\ast i}(t,\textbf{v}) & - \mathfrak{s}_{1}^i(t,\textbf{v})- \mathfrak{s}_{2}^i(t,\textbf{v}) \end{pmatrix}.                                                                                       
\end{align}
The neutrino density $g_i$ is assumed to be independent of time and space and has to fulfill the normalisation condition $\sum_{i=1}^N|g_i|=2$.  Note that $\mathfrak{s}_1^i$ and $\mathfrak{s}_2^j$ have to be real for all modes and that in general $\mathcal{S}_{\alpha\beta}^i\in\mathbb{C}$.  
In this paper, our focus is on flavor conversion, and hence we are interested in the off-diagonal entries of the density matrix. The differential equation we aim to solve is thus
\begin{align}
\renewcommand*{\arraystretch}{1.5}
i\partial_t\: \frac{g_i}{3}
 \begin{pmatrix}
  \mathcal{S}_{ex}^i(t,\textbf{v})\\
  \mathcal{S}_{ey}^i(t,\textbf{v})\\
  \mathcal{S}_{xy}^i(t,\textbf{v})
 \end{pmatrix}
=
\underbrace{
\begin{pmatrix}
 \left[H_{E_i}^{\text{vac}},\rho_i\right]_{ex}\\
  \left[H_{E_i}^{\text{vac}},\rho_i\right]_{ey}\\
 \left[H_{E_i}^{\text{vac}},\rho_i\right]_{xy}
\end{pmatrix}}_{\textrm{(I)}}
+
\underbrace{
\begin{pmatrix}
 \left[H^{\lambda},\rho_i\right]_{ex}\\
  \left[H^{\lambda},\rho_i\right]_{ey}\\
  \left[H^{\lambda},\rho_i\right]_{xy}
\end{pmatrix}}_{\textrm{(II)}}
+
\underbrace{
\begin{pmatrix}
 \left[H^{\nu\nu}(\textbf{v}),\rho_i\right]_{ex}\\
  \left[H^{\nu\nu}(\textbf{v}),\rho_i\right]_{ey}\\
  \left[H^{\nu\nu}(\textbf{v}),\rho_i\right]_{xy}
\end{pmatrix}}_{\textrm{(III)}},
\label{eq:Goal}
\end{align}
with the initial conditions $\mathfrak{s}^i_1(0,\textbf{v})=2$, $\mathfrak{s}^i_2(0,\textbf{v})=-1$ and $\mathcal{S}^i_{\alpha\beta}(0,\textbf{v})=0$, assuming that all neutrinos are produced as electron neutrinos and antineutrinos.
\paragraph{Vacuum-Term (I):}
To obtain the first term of \eref{eq:Goal}, we calculate the transformed mass matrix and the commutator with the density matrix of mode $i$. It is useful to introduce the oscillation frequency $\omega_i:=\Delta m_{21}^2/(2E_i)$ and the mass-squared difference ratio $\eta:=\Delta m_{31}^2/\Delta m_{21}^2$ and order the terms with respect to the latter one. Solving the commutator yields
\begin{align}
\renewcommand*{\arraystretch}{1.5}
\frac{3}{g_i\omega_i}
\textbf{\textrm{(I)}}
=\left[
 \underline{\underline{A_0}}
+
 \underline{\underline{B_0}}\eta\right]
  \begin{pmatrix}
  \mathcal{S}_{ex}^i\\
  \mathcal{S}_{ey}^i\\
  \mathcal{S}_{xy}^i
 \end{pmatrix}
+\left[\underline{\underline{A_1}}+\underline{\underline{B_1}}\eta\right]
\begin{pmatrix}
 \mathfrak{s}_{1}\\
 \mathfrak{s}_{2}\\
 -(\mathfrak{s}_1+\mathfrak{s}_2)
\end{pmatrix}+\left[\underline{\underline{A_2}}+\underline{\underline{B_2}}\eta\right]
\begin{pmatrix}
   \mathcal{S}_{ex}^{\ast i}\\
  \mathcal{S}_{ey}^{\ast i}\\
  \mathcal{S}_{xy}^{\ast i}.
\end{pmatrix}.
\label{eq:VacTermDec}
\end{align}
The matrices $\underline{\underline{A_n}}$ and $\underline{\underline{B_n}}$ contain combinations of the mixing angles and read
\begin{gather}
 \underline{\underline{A_0}}:=
 \begin{pmatrix}
  -c_{12}^2+c_{13}^2s_{12}^2 & \frac{1}{2}S_{12}s_{13} &0\\
  \frac{1}{2}S_{12}s_{13} & C_{13} s_{12}^2 & \frac{1}{2}S_{12} c_{13} \\
0 & \frac{1}{2}S_{12} c_{13} & c_{12}^2-s_{12}^2s_{13}^2
 \end{pmatrix},
\quad
 \underline{\underline{B_0}}:=
 \begin{pmatrix}
  s_{13}^2 & 0 & 0\\
  0 & -C_{13} & 0\\
 0 & 0 & -c_{13}^2
 \end{pmatrix},\nonumber\\[0.5cm]
 \underline{\underline{A_1}}:=\begin{pmatrix}
  -\frac{1}{2}S_{12}c_{13} & \frac{1}{2}S_{12}c_{13} & 0\\
  S_{13}s_{12}^2 & \frac{1}{2}S_{13}s_{12}^2 & 0\\
  \frac{1}{2}S_{12}s_{13} & S_{12}s_{13}& 0
 \end{pmatrix}
,\quad
 \underline{\underline{B_1}}:=\begin{pmatrix}
       0 & 0 &0 \\
       -S_{13} & -\frac{1}{2}S_{13} & 0 \\
       0 & 0 & 0
      \end{pmatrix},\\[0.5cm]
 \underline{\underline{A_2}}:=\begin{pmatrix}
       0 & 0 & -\frac{1}{2}S_{13}s_{12}^2\\
       0 & 0 & 0 \\
       \frac{1}{2}S_{13}s_{12}^2 & 0 & 0
      \end{pmatrix}
,\quad
 \underline{\underline{B_2}}:=\begin{pmatrix}
       0 & 0 & \frac{1}{2}S_{13}\\
       0 & 0& 0\\
       -\frac{1}{2}S_{13} & 0 & 0
      \end{pmatrix},\nonumber
\end{gather}
where we abbreviate $c_{kl}:=\cos(\theta_{kl})$, $s_{kl}:=\sin(\theta_{kl})$ for mixing angle $\theta_{kl}$ and refer by capital $S_{kl}$ to $\sin(2\theta_{kl})$. Here, the introduced double underline notation refers to matrices that act on vectors spanned by the off-diagonal density matrix elements. We will use this notation throughout the paper from now on.

\paragraph{Matter-Term (II):}
The matter-term describes the interaction between the neutrino ensemble and the surrounding matter. These interactions are dominated by charged current processes of electrons and electron-neutrinos while the the densities of muons and tau leptons are very small. By defining the effective MSW-potential strength  $\lambda:=\sqrt{2}G_F N_e$, the Hamiltonian for the matter effect becomes
\begin{align}
H^{\lambda}=
 \begin{pmatrix}
  \lambda & 0 & 0\\
  0 &0 & 0\\
  0 & 0 &0
 \end{pmatrix}.
\end{align}
Calculating the commutator with the density matrix and reading off the upper off-diagonal components we find 
\begin{align}
\label{eq:MatterTerm}
\frac{3}{g_i}\textbf{\textrm{(II)}}=
 \begin{pmatrix}
  \lambda & 0 & 0\\
  0 & \lambda & 0 \\
  0 & 0 & 0
 \end{pmatrix}
  \begin{pmatrix}
  \mathcal{S}_{ex}^i(t,\textbf{v})\\
  \mathcal{S}_{ey}^i(t,\textbf{v})\\
  \mathcal{S}_{xy}^i(t,\textbf{v})
 \end{pmatrix}.
\end{align}

\paragraph{Neutral-Current-Term (III):} 
The neutrino-neutrino interaction term includes the commutator of two density matrices of different modes which leads to higher order terms in the flavor cross correlation terms. We omit terms of higher than linear order in the flavor cross correlations $|\mathcal{S}_i|^2\ll 1$ which  implies that the flavor densities will not depart a lot from the initial value $\mathfrak{s}(t,\textbf{v})\approx \mathfrak{s}(0,\textbf{v}) $ $\forall t$ . We find  
\begin{align}
 \left[\rho_j,\rho_i\right]\approx\frac{g_ig_j}{9}
 \begin{pmatrix}
  0 & \mathfrak{s}_{1}^{[j,}\mathcal{S}_{ex}^{i]}+\mathcal{S}_{ex}^{[j,}\mathfrak{s}_{2}^{i]} & \mathfrak{s}_{1}^{[j,}\mathcal{S}_{ey}^{i]}+\mathcal{S}_{ey}^{[i,}(\mathfrak{s}_{1}+\mathfrak{s}_{2})^{j]}  \\
  \mathcal{S}_{ex}^{\ast[j,}\mathfrak{s}_{1}^{i]}+\mathfrak{s}_{2}^{[j,}(\mathcal{S}_{ex}^{\ast})^{i]} & 0 & \mathfrak{s}_{2}^{[j,}\mathcal{S}_{xy}^{i]}+\mathcal{S}_{xy}^{[i,}(\mathfrak{s}_1+\mathfrak{s}_2)^{j]}\\
  \mathfrak{s}_1^{[j,}(\mathcal{S}_{ey}^{\ast})^{i]}+\mathcal{S}_{ey}^{\ast[j,}(\mathfrak{s}_1+\mathfrak{s}_2)^{i]} & \mathcal{S}_{xy}^{\ast[j,}\mathfrak{s}_2^{i]}+\mathcal{S}_{xy}^{\ast[j,}(\mathfrak{s}_1+\mathfrak{s}_2)^{i]} & 0
 \end{pmatrix},
\end{align}
where we have introduced the following notation
\begin{align}
 \mathcal{S}_a^{[i,}\mathfrak{s}_b^{j]}:=\mathcal{S}_a^i\mathfrak{s}_b^j-\mathcal{S}_b^j\mathfrak{s}_a^i.
\end{align}

We conclude that the contributions to the equations for the off-diagonal matrix elements of the density matrix are
\begin{align}
\frac{3}{g_i}\textbf{\textrm{(III)}}=
 \mu\sum_{j=1}^N\frac{g_j}{3}(1-\textbf{v}^i\textbf{v}^j)
            \begin{pmatrix}
            \mathfrak{s}_{1}^{[j,}\mathcal{S}_{ex}^{i]}+\mathcal{S}_{ex}^{[j,}\mathfrak{s}_{2}^{i]}\\
            \mathfrak{s}_{1}^{[j,}\mathcal{S}_{ey}^{i]}+\mathcal{S}_{ey}^{[i,}(\mathfrak{s}_{1}+\mathfrak{s}_{2})^{j]} \\
              \mathfrak{s}_{2}^{[j,}\mathcal{S}_{xy}^{i]}+\mathcal{S}_{xy}^{[i,}(\mathfrak{s}_1+\mathfrak{s}_2)^{j]}
            \end{pmatrix}.\centering
\label{eq:long}
\end{align}

\subsection{Eigenvalue equation}
Having introduced the theoretical description of flavor conversion in a dense environment and derived a matrix equation for the flavor correlations, we present an approximated solution for this equation in this section. 

The flavor correlations are often treated as plane waves $\mathcal{S}_i(t,\textbf{v})=Q_i(\textbf{v})e^{-i\Omega t}$. Flavor conversion occurs if the frequency $\Omega$ has an imaginary part that leads to an exponentially growing flavor correlation function. Our main focus shall therefore be to find those instabilities. A consequence of this behavior is that we can neglect the constant term of \eref{eq:VacTermDec}.

Due to the environmental conditions in a SN, we assume from now on a \textit{large matter potential}. This leads to a suppression of off-diagonal terms in \eref{eq:VacTermDec}, and we hence neglect the third term of the equation.\footnote{We have also included the third term of \eref{eq:VacTermDec} under the assumption that $S$ is purely imaginary or real without changing the results.} We will justify this more carefully in \sref{sec:PropStates}.

With these simplifications and the plane wave ansatz, \eref{eq:Goal} reduces to an eigenvalue equation for the frequency $\Omega$
\begin{equation}
\begin{aligned}
 \Omega 
 \begin{pmatrix}
   Q_{ex}^i\\
   Q_{ey}^i\\
   Q_{xy}^i
 \end{pmatrix}
 =&\left[
 \underline{\underline{A_0}}\omega_i+\underline{\underline{B_0}}\eta\omega_i+
   \lambda\begin{pmatrix}
  1 & 0 & 0\\
  0 & 1 & 0 \\
  0 & 0 & 0
 \end{pmatrix}\right]
  \begin{pmatrix}
   Q_{ex}^i\\
   Q_{ey}^i\\
   Q_{xy}^i
 \end{pmatrix}\\
 &+  \mu\sum_{j=1}^Ng_j(1-\textbf{v}^i\textbf{v}^j)
            \begin{pmatrix}
             (Q_{ex}^i-Q_{ex}^j)\\
             (Q_{ey}^i-Q_{ey}^j)\\
             0
            \end{pmatrix}.\centering
\end{aligned}
\end{equation}
The density of the $i$th mode $g_i/3$ appears in every term and hence cancels out. The normalization factor of $1/3$ of $g_j$ cancels a factor of $3$ arising from the initial conditions of $\mathfrak{s}_{1,2}^{i,j}$ which we assumed to approximately constant in time. By solving this equation, it can be determined if $\Omega$ has an imaginary part and linearly unstable solution thus exist.

\section{Toy systems}
\label{sec:models}
In order to investigate and explore the developed model, we consider the case of two colliding and four intersecting neutrino beams. These pictures were already used in the literature \cite{Chakraborty:2016lct,Raffelt:2013isa,Mangano:2014zda,Hansen:2014paa,Duan:2014gfa,Mirizzi:2015fva,Chakraborty:2015tfa,Abbar:2015mca} and allow us to compare our results with the two-neutrino limit. For the mixing angles we take the following values  \cite{Esteban:2016qun}
\begin{align}
 \theta_{12}= 33.62^{\circ},\qquad
 \theta_{23}= 47.2^{\circ},\qquad
 \theta_{13}= 8.54^{\circ}.
\end{align}
We checked that our results are insensitive to changes within the uncertainties.

\subsection{Two-Beam-System}
The simplest model to study is a system of two different neutrino modes that propagate in one dimension and can be interpreted as two colliding beams. Let us assume a left moving antineutrino beam and a right moving neutrino beam denoted by
\begin{align}\underline{Q}^{\bar{l}}=
 \begin{pmatrix}
    \bar{L}_{ex}\\
  \bar{L}_{ey}\\
  \bar{L}_{xy}
 \end{pmatrix}
\quad\textrm{and}\quad
\underline{Q}^{r}=
\begin{pmatrix}
  R_{ex}\\
 R_{ey}\\
 R_{xy}
 \end{pmatrix}.
\end{align}
We collect them in a six dimensional vector $\underline{Q}:=(\underline{Q}^{r},\underline{Q}^{\bar{l}})^T$.  
Parameters referring to the left moving antineutrino beam pick up an index $\bar{l}$ while parameters belonging to the right moving beam are labeled by an index $r$.
For the antineutrino beam, we have $\omega\to-|\omega|$ due to our convention, and the neutral-current term $(1-\textbf{v}_i\textbf{v}_j)$ becomes $0$ for $i=j$ (parallel moving modes) and $2$ for $i\neq j$ (oppositely moving modes). This leads to a $6\times6$ dimensional matrix which we write as a matrix of $3\times3$ block matrices.  For equal numbers of neutrinos and antineutrinos $-g_{\bar{l}}=g_r=1$, the equation reduces to
\begin{align}
 \Omega\, \underline{Q}=\left[
 \begin{pmatrix}
 |\omega|\underline{\underline{A_0}}+|\omega|\eta\underline{\underline{B_0}}+(\lambda-2\mu)\underline{\underline{\Lambda}} & +2\mu\underline{\underline{\Lambda}}\\
 -2\mu\underline{\underline{\Lambda}} & -(|\omega|\underline{\underline{A_0}}+|\omega|\eta\underline{\underline{B_0}}-(\lambda+2\mu)\underline{\underline{\Lambda}})
\end{pmatrix}\right]
\underline{Q},
\label{eq:2BE}
\end{align}
where $\underline{\underline{\Lambda}}$ is defined as
\begin{align}
\underline{\underline{\Lambda}}:=
\begin{pmatrix}
 1 & 0 & 0\\
 0 & 1 &0 \\
 0& 0 & 0
\end{pmatrix}.
\end{align}
Since we have a $6\times 6$ matrix, we cannot solve the equation analytically, but have to resolve to a numerical solution.

\subsubsection{Results}

\begin{figure}[tbp]
 \centering
  \includegraphics[width=1\textwidth]{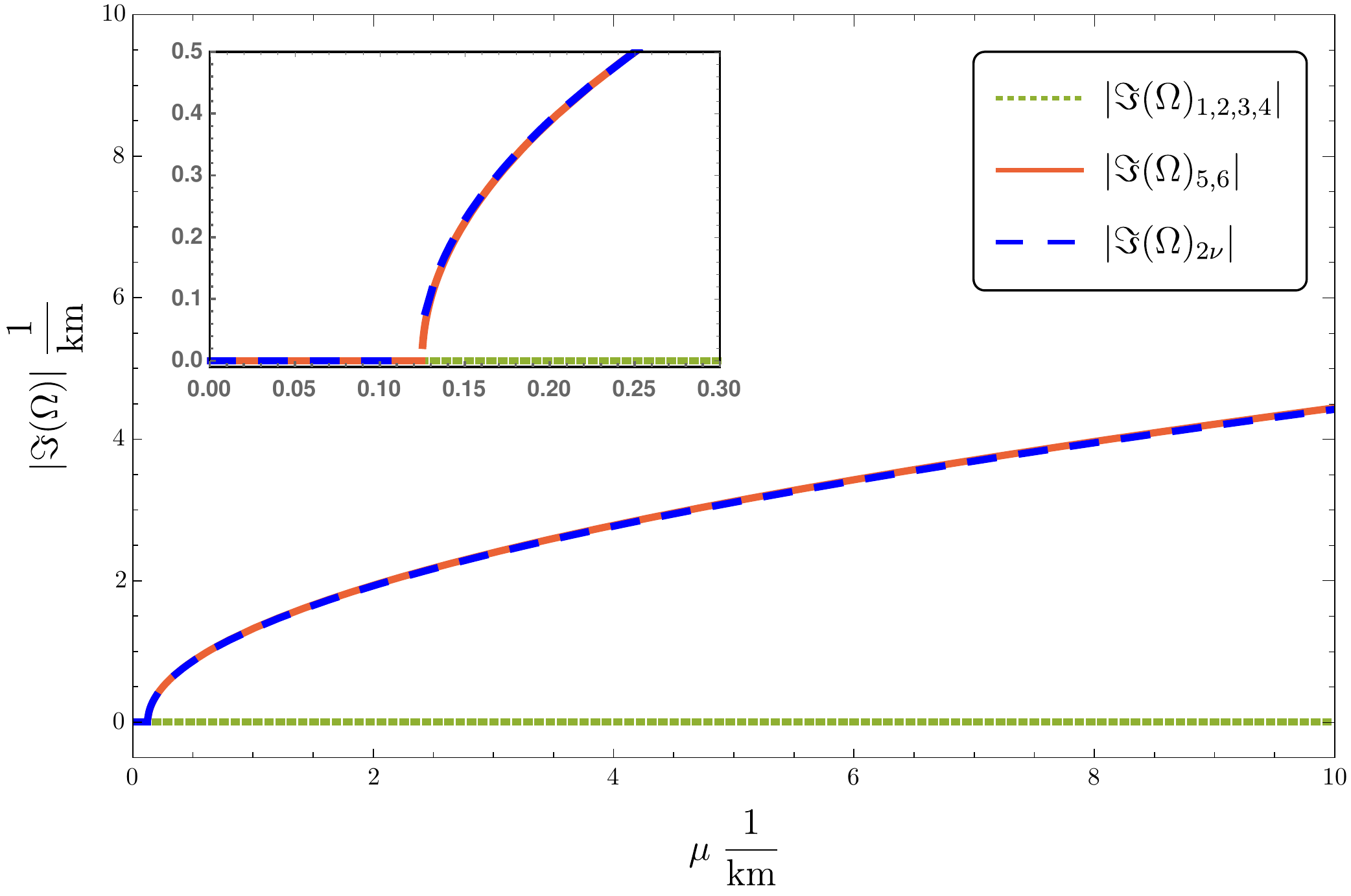}
  \caption{Imaginary parts of all eigenvalues for the two beam model as a function of the neutrino potential in the two-neutrino limit with IO (green and orange lines) compared with \eref{eq:2BE} (blue dashed line). }
  \label{fig:2NLimitIH}
\end{figure}
Before exploring new features of the matrix in \eref{eq:2BE}, we convince ourselves that in the two-neutrino limit ($\theta_{13}\approx 0^{\circ}$), we are able to reproduce the result found in \cite{Chakraborty:2016lct} 
\begin{align}
 \Omega_{2\nu}=\sqrt{\frac{\Delta m_{31}^2}{2E_i}\cdot\left(\frac{\Delta m_{31}^2}{2E_i}+4\mu\right)}.
 \label{eq:2NLR}
\end{align}
In \fref{fig:2NLimitIH} we show for IO the imaginary parts of the eigenvalues of \eref{eq:2BE} in the two-neutrino limit ($\theta_{13}=0$) as orange lines as a function of $\mu$.  The  two-neutrino result in \eref{eq:2NLR} is shown by the blue dashed curve which agrees very well with our result. For NO all eigenvalues $\Omega$ are real and thus no unstable modes exist.

\begin{figure}[tbp]
 \centering
  \includegraphics[width=1\textwidth]{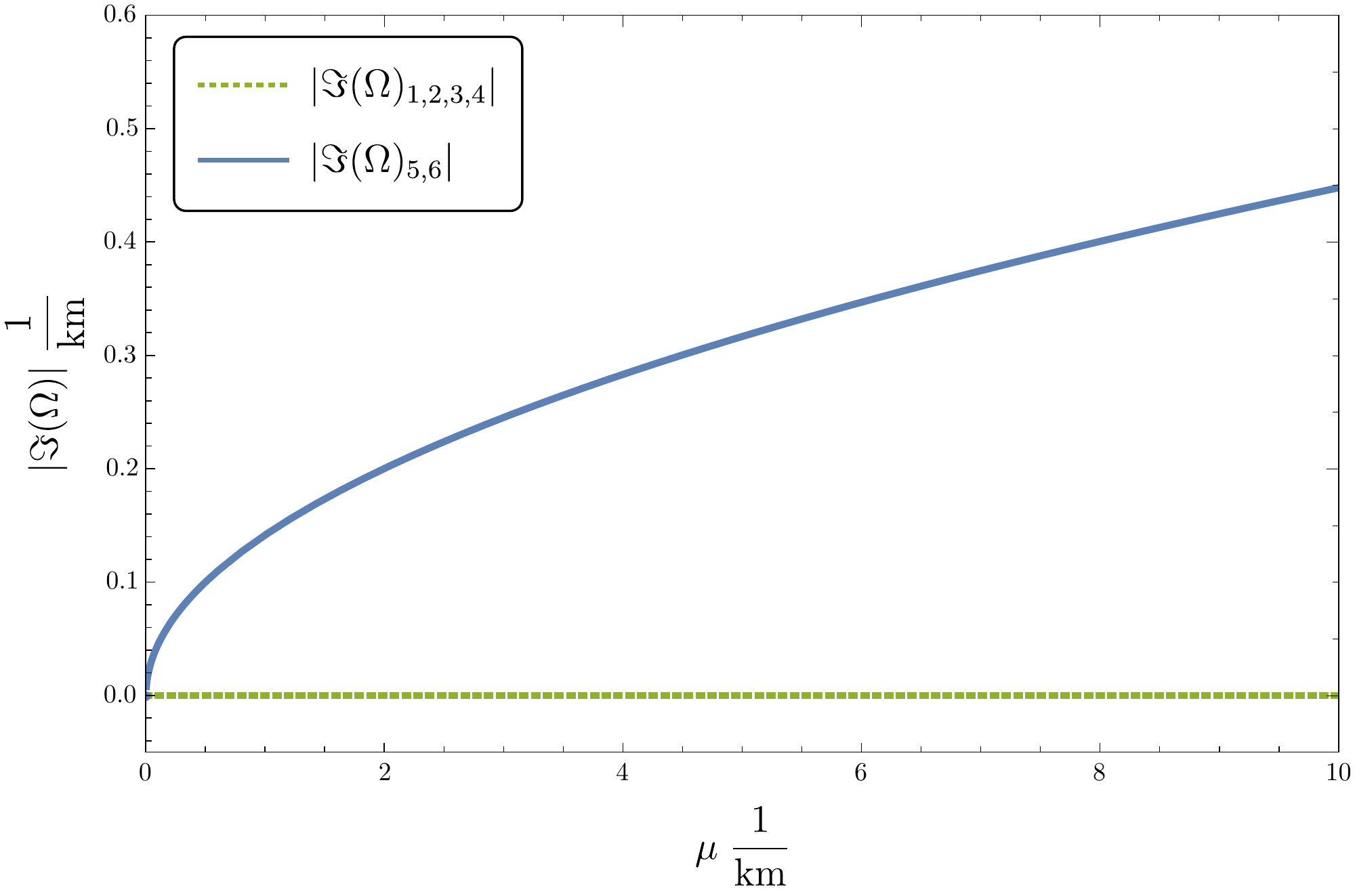}
  \caption{Imaginary parts of all eigenvalues for the two beam model as a function of the neutrino potential with the full three flavor equation for NO.}
  \label{fig:3NNO}
\end{figure}

We now return to the case of three neutrino flavors and calculate numerically the eigenvalues of the matrix \eref{eq:2BE}. In the presence of a large matter potential of $\lambda=100\,\textrm{km}^{-1}$, a mass square ratio $|\eta|=33$, and an oscillation frequency $\omega=0.015\,\textrm{km}^{-1}$, we find results identical to the two-neutrino limit for IO. For NO, however, unstable modes are present in the three neutrino case with $\theta_{13}\neq 0$. Comparing \fref{fig:3NNO} and \fref{fig:2NLimitIH} also shows, that the growth rate for the new instability in NO is smaller than for the instability in the case of IO. The instability for NO depends on the mixing angles $\theta_{13}$, $\theta_{23}$ as well as on the mass square ratio $\eta$. Numerically, we also confirm that the rotation $R_{23}$ does not change the results as expected.

\subsection{Four-Beam-System}
\begin{figure}[tbp]
 \centering
  \includegraphics[width=0.4\textwidth]{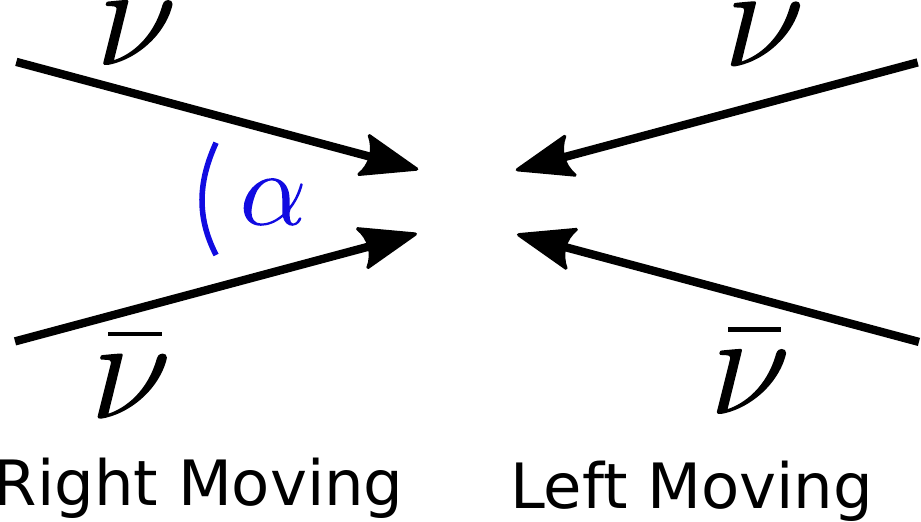}
  \caption{Geometry of the four-beam model.}
  \label{fig:4BSketch}
\end{figure}

Our next step is to generalize the two beam picture to a four beam configuration in which an additional free parameter, the intersection angle $\alpha$, is introduced. We consider a configuration in which a right moving neutrino beam and a left moving antineutrino beam separated by the relative angle interact with a left moving neutrino beam and a left moving antineutrino beam as it is shown in figure \fref{fig:4BSketch}. Our goal is to investigate whether the instability we have found in the two beam configuration for NO is also present in this model. 
In the limit of zero relative angle and zero left moving antineutrino and right moving neutrino density, the four beam system coincides with the two beam system discussed before.

The generalization of \eref{eq:2BE} to the case of four intersecting beams with relative angle $\alpha$ is 

\begin{align}
 \Omega\,\underline{Q}=&\left\{|\omega|
 \begin{pmatrix}
  \underline{ \underline{A}}+\eta\underline{\underline{B}} & 0 & 0 & 0 \\
  0 & -\underline{\underline{A}}-\eta\underline{\underline{B}} & 0 & 0\\
    0&0&\underline{ \underline{A}}+\eta\underline{\underline{B}} & 0 \\
  0&0&0 & -\underline{\underline{A}}-\eta\underline{\underline{B}}
 \end{pmatrix}\right.
 +\lambda
 \begin{pmatrix}
  \underline{\underline{\Lambda}} & 0 & 0 & 0\\
  0 & \underline{\underline{\Lambda}}& 0 & 0\\
  0 & 0&\underline{\underline{\Lambda}} & 0\\
  0 & 0&0 & \underline{\underline{\Lambda}}
 \end{pmatrix}\nonumber\\
&\quad+\mu
 \left[2\begin{pmatrix}
  g_{\bar{l}}\underline{\underline{\Lambda}} & 0 & 0 & -g_{\bar{l}}\underline{\underline{\Lambda}} \\
    0 &   g_{l} \underline{\underline{\Lambda}} & -g_l\underline{\underline{\Lambda}} & 0\\
     0 & -g_{\bar{r}}\underline{\underline{\Lambda}} & g_{\bar{r}}\underline{\underline{\Lambda}} &  0 \\
    -g_r\underline{\underline{\Lambda}} &0 & 0&   g_{r} \underline{\underline{\Lambda}} 
 \end{pmatrix}\right.\\
&\quad+(1-\cos\alpha)
\begin{pmatrix}
  g_{\bar{r}} \underline{\underline{\Lambda}} & -g_{\bar{r}}  \underline{\underline{\Lambda}}& 0 & 0 \\
   -g_{r}  \underline{\underline{\Lambda}} &   g_{r}\underline{\underline{\Lambda}}& 0 & 0\\
   0 & 0 &g_{\bar{l}} \underline{\underline{\Lambda}} & -g_{\bar{l}}  \underline{\underline{\Lambda}} \\
   0 & 0 &-g_{l}  \underline{\underline{\Lambda}} &   g_{l}\underline{\underline{\Lambda}}
\end{pmatrix}\nonumber\\
 &\quad+(1+\cos\alpha)\left.\left.
 \begin{pmatrix}
  g_l   \underline{\underline{\Lambda}}& 0 &-g_l  \underline{\underline{\Lambda}} & 0\\
 0 &g_{\bar{l}}  \underline{\underline{\Lambda}} & 0 & -g_{\bar{l}} \underline{\underline{\Lambda}}\\
  -g_r  \underline{\underline{\Lambda}} & 0& g_r   \underline{\underline{\Lambda}}& 0\\
  0& -g_{\bar{r}} \underline{\underline{\Lambda}} & 0 &g_{\bar{r}}  \underline{\underline{\Lambda}}
 \end{pmatrix}\right]\right\}\underline{Q},\nonumber
\end{align}
where $\underline{Q}:=(\underline{R},\underline{\bar{R}},\underline{L},\underline{\bar{L}})^T$ is a twelve dimensional vector with $\underline{R}:=(R_{ex},R_{ey},R_{xy})^T$ and similarly for the other amplitudes. The neutrino-neutrino term can be decomposed in a symmetric and antisymmetric term with respect to the propagation direction of the mode.
We perform a basis transformation $A_{\pm}:=\frac{1}{2}(L\pm R)$ and $\bar{A}_{\pm}:=\frac{1}{2}(\bar{L}\pm\bar{R})$ which can be written in matrix form
\begin{align}
 \begin{pmatrix}
  \underline{A}_+\\
  \underline{\bar{A}}_+\\
  \underline{A}_-\\
  \underline{\bar{A}}_-
 \end{pmatrix}:=
  \frac{1}{2}
 \begin{pmatrix}
  \mathbbm{1}_3 & 0 & \mathbbm{1}_3 & 0\\
  0 & \mathbbm{1}_3 & 0 & \mathbbm{1}_3\\
  -\mathbbm{1}_3 & 0 & \mathbbm{1}_3 & 0\\
  0 & -\mathbbm{1}_3 & 0 & \mathbbm{1}_3
 \end{pmatrix}
  \begin{pmatrix}
  \underline{R}\\
  \underline{\bar{R}}\\
  \underline{L}\\
  \underline{\bar{L}}
 \end{pmatrix}.
\end{align}
While this transformation leaves the vacuum- and matter terms invariant, it changes the three contributions to the neutrino-neutrino term. Setting the neutrino and antineutrino densities to $g_{r}=g_{l}=\frac{1}{2}$ and $g_{\overline{r}}=g_{\overline{l}}=-\frac{1}{2}$, the neutral current term becomes
\begin{align}\mu\left[
 \begin{pmatrix}
    -\underline{\underline{\Lambda}}&\underline{\underline{\Lambda}}&0 & 0\\
  -\underline{\underline{\Lambda}} & \underline{\underline{\Lambda}} & 0& 0 \\
    0&0&-\underline{\underline{\Lambda}} & -\underline{\underline{\Lambda}}\\
  0 & 0 & \underline{\underline{\Lambda}}& \underline{\underline{\Lambda}} 
 \end{pmatrix}+
 \frac{(1-\cos\alpha)}{2}
 \begin{pmatrix}
 -\underline{\underline{\Lambda}} &   \underline{\underline{\Lambda}} &   0& 0\\
    -\underline{\underline{\Lambda}} &  \underline{\underline{\Lambda}}&  0 & 0\\
    0 &  0 & -\underline{\underline{\Lambda}} &  \underline{\underline{\Lambda}}\\
   0 &   0 &   -\underline{\underline{\Lambda}}& \underline{\underline{\Lambda}}\\
 \end{pmatrix}+
 (1+\cos\alpha)
 \begin{pmatrix}
 0&0 & 0 & 0\\
  0&0&   0 &  0\\
   0 & 0 & \underline{\underline{\Lambda}} &  0\\
    0 &  0 &0 &-\underline{\underline{\Lambda}}\\
 \end{pmatrix}\right],
\end{align}
The transformation separates the equations into two decoupled blocks corresponding to the symmetric and the antisymmetric amplitudes.
In the cases of $\cos\alpha=1$ and $\cos\alpha=-1$, we recover with the upper block of equations the two beam model.

\subsubsection{Results}

\begin{figure}[tbp]
 \centering
  \includegraphics[width=0.8\textwidth]{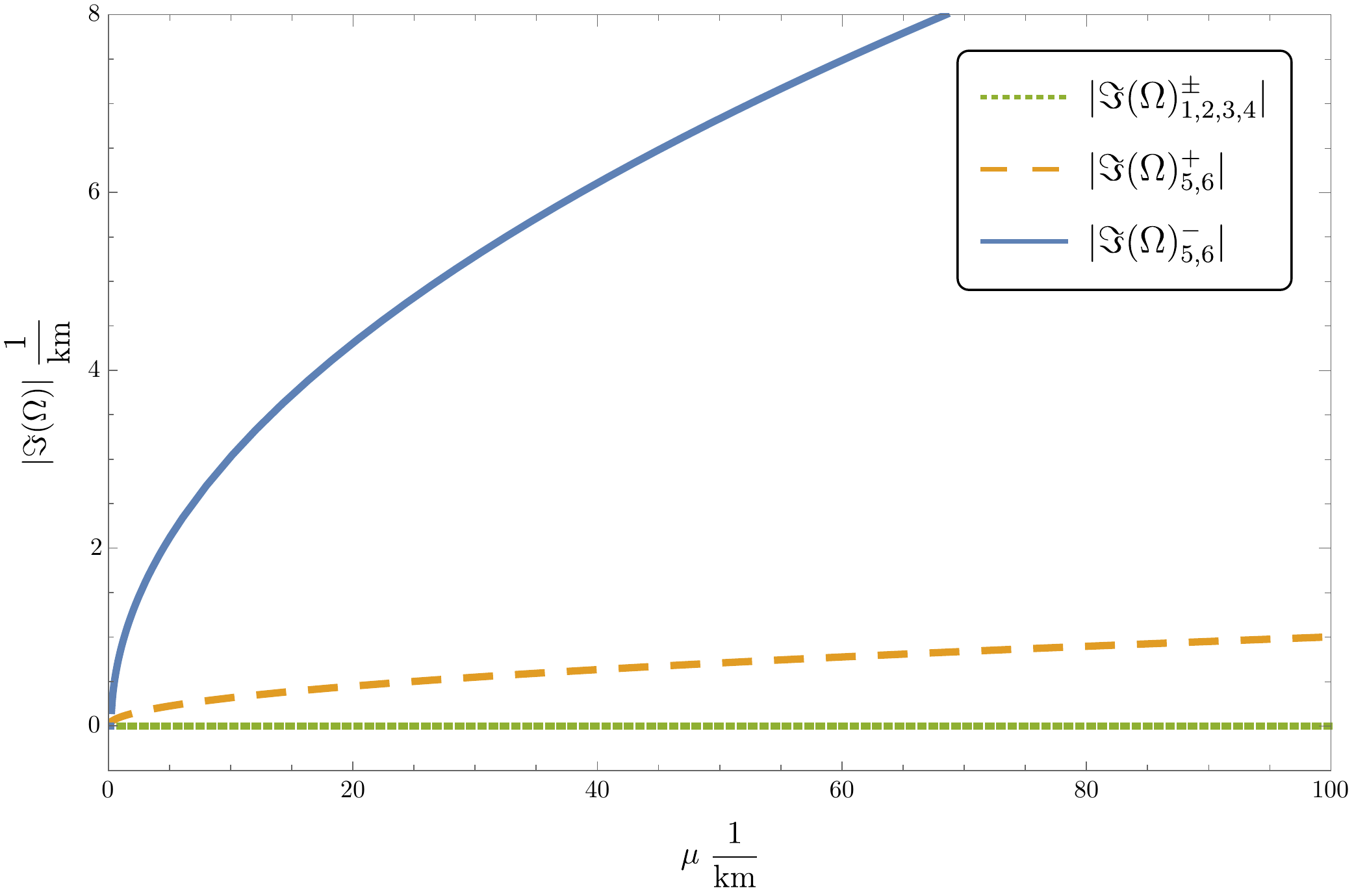}
  \caption{Instabilities in the four beam configuration with relative angle $\alpha=0^{\circ}$ for NO. There exist growing instabilities in both, asymmetric and symmetric modes, but the latter ones grow siginificantly slower than the former ones.}
  \label{fig:3NNOTest}
\end{figure}

\begin{figure}[tbp]
   \centering
   \includegraphics[width=0.47\textwidth]{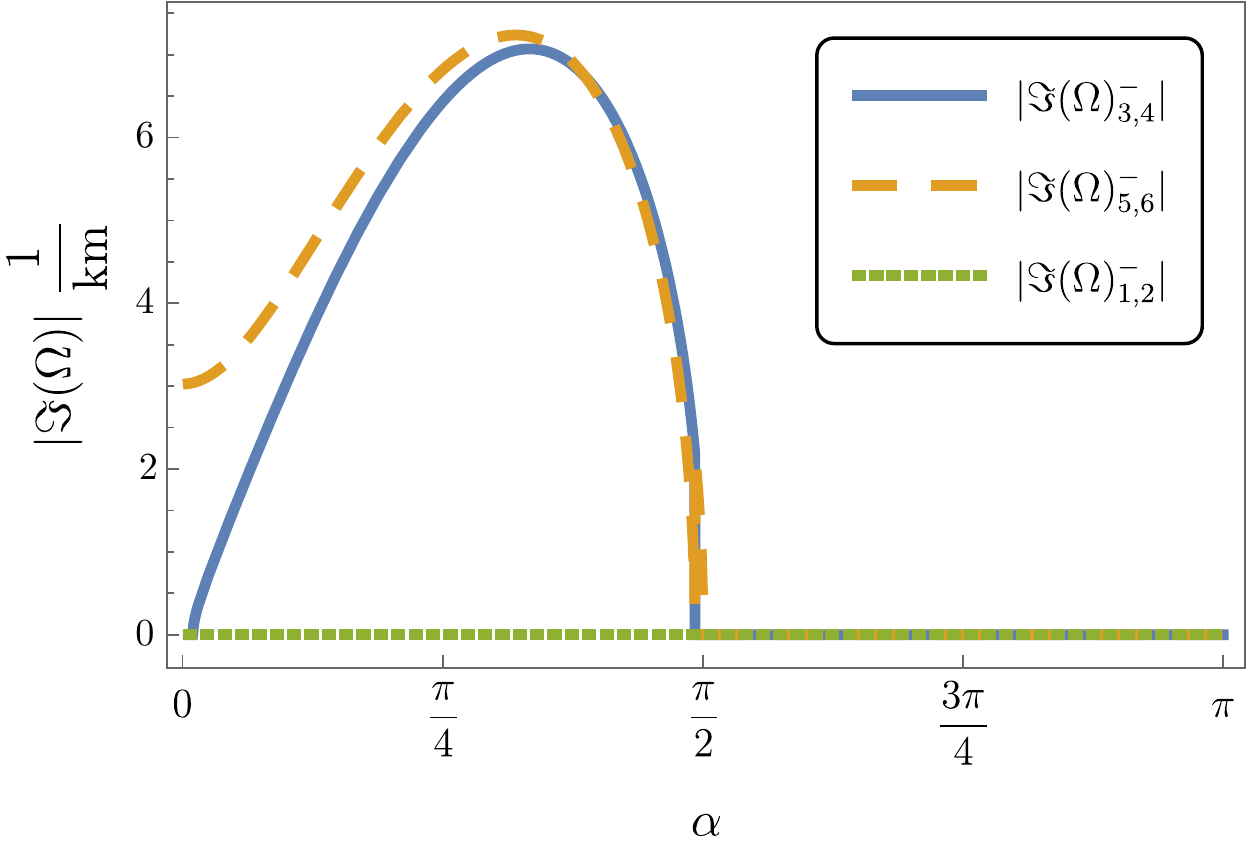}
   \includegraphics[width=0.47\textwidth]{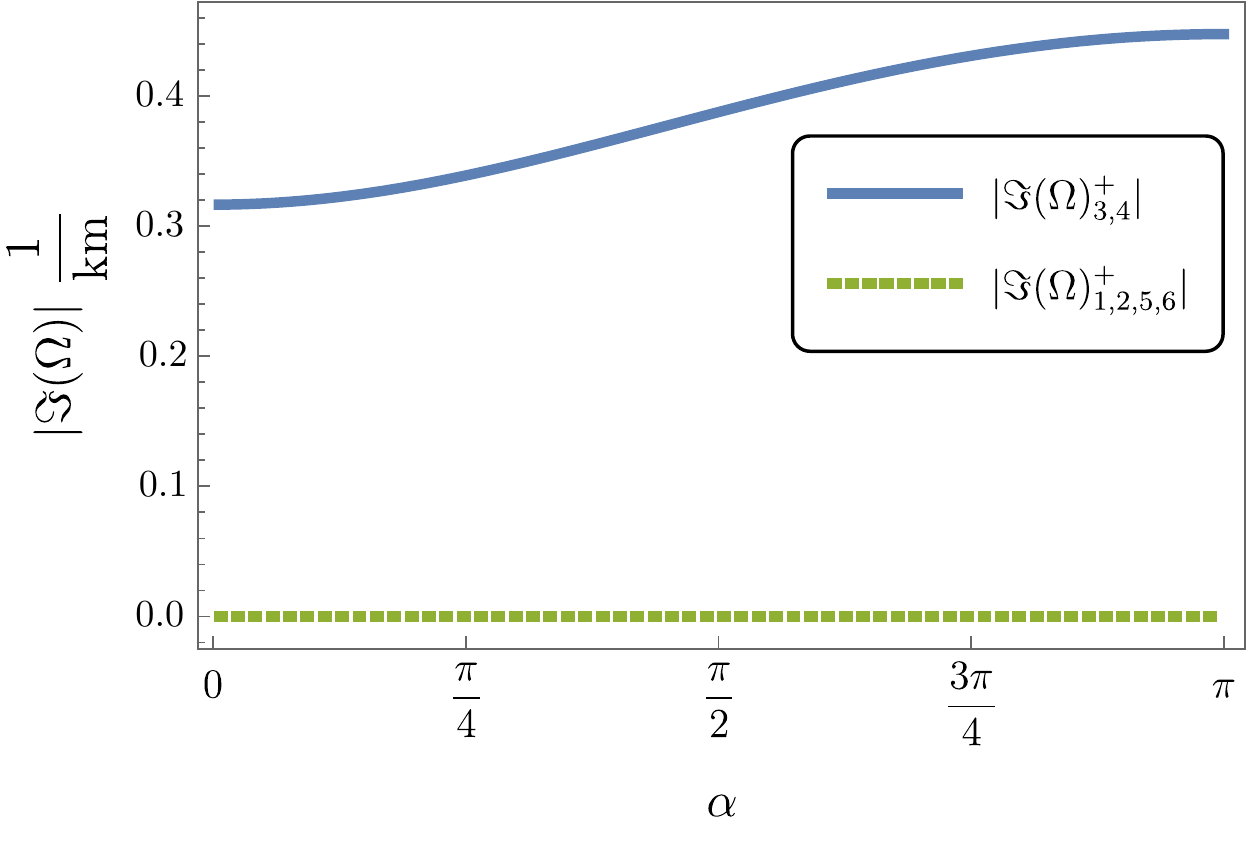}
   \caption{Angle dependencies of modes in the four beam model for NO and for $\mu=10\,\frac{1}{\textrm{km}}$. \emph{Left:} Angle dependence of the imaginary part of the antisymmetric eigenvalues in the four beam model for NO. \emph{Right:} Angle dependence of the imaginary part of the symmetric eigenvalues in the four beam model for NO.}
   \label{fig:2NLimitIHg}
\end{figure}

We solve the eigenvalue equation again numerically with the same parameters as in the previous section. Next to fast growing instabilities for the asymmetric modes, we also find instabilities for the symmetric modes with significantly smaller growth rates, see \fref{fig:3NNOTest}. As for the two beam model, the instabilities for the symmetric modes depend on both mixing angles and the square mass ratio $\eta$. In \fref{fig:2NLimitIHg} we show the behavior of the eigenvalues as a function of the intersection angle $\alpha$. The fast growing antisymmetric instability vanishes for an intersection angle of $\alpha\ge\pi/2$ while the imaginary part of an eigenvalue belonging to a symmetric mode still grows.

\section{Analysis in the basis of propagating states}
\label{sec:PropStates}
Until now, we have performed the calculations in the $(\nu_e, \nu_x, \nu_y)$-basis. However, as it is noted in \cite{Hansen:2019}, it is beneficial to perform the stability analysis in the basis of propagating states since a diagonal density matrix here correspond to a 'fixed point'.
Under the assumption of adiabatic evolution, the basis of propagating states is coinciding with the basis of instantaneous eigenstates in which the Hamiltonian is diagonal.

For understanding our results, it turns out that we can ignore the $H^{\nu\nu}$ term when determining the basis of propagating states. Then the Hamiltonian can be written as
\begin{equation}
  \label{eq:Hflavour}
      H = \omega_i 
  \begin{pmatrix}
    s_{12}^2c_{13}^2 & s_{12} c_{13} c_{12} & -s_{12}^2 c_{13} s_{13}\\
    s_{12}c_{12}c_{13} & c_{12}^2 & s_{13} s_{12} c_{12} \\
    -s_{12}^2 c_{13} s_{13} & s_{13} c_{12} s_{12} & s_{12}^2 s_{13}^2 
  \end{pmatrix}
  + \omega_i \eta
  \begin{pmatrix}
    s_{13}^2 & 0 & s_{13} c_{13} \\
    0 & 0 & 0\\
    s_{13} c_{13} & 0 & c_{13}^2
  \end{pmatrix}  +
  \begin{pmatrix}
    \lambda & 0 & 0\\ 0 & 0 & 0\\ 0 & 0 & 0
  \end{pmatrix} .
\end{equation}
in the ($\nu_e,\nu_x,\nu_y$)-basis.

In the limit of $\lambda \gg \omega \eta$, the diagonalised Hamiltonian is 
\begin{equation}
  \label{eq:Hdiag}
  H'_{\rm diag} \approx
  \begin{pmatrix}
    \lambda + \omega_i (s_{12}^2c_{13}^2 + \eta s_{13}^2) & 0 & 0 \\
    0  & \omega_i c_{12}^2 & 0 \\
    0 & 0 & \omega_i ( s_{12}^2 s_{13}^2 + \eta c_{13}^2)
  \end{pmatrix}
\end{equation}
up to terms of order $(s_{13}c_{13}\omega \eta)^2/\lambda$. This approximation basically corresponds to neglecting the off-diagonal terms of $H^{\rm vac}_{i}$, and for two-neutrino oscillations it corresponds to the statement that $\omega_{\rm eff} \approx \lambda - c_{2\theta} \omega$ \cite{Hansen:2019}. The result in \eref{eq:Hdiag} agree with the diagonal matrix derived in~\cite{Duan:2008za} in the limit $\theta_{13}=0$, but we are more interested in non-zero $\theta_{13}$. When going to the basis of propagating states, we also transform $\rho_i$ to $\rho'_i$:
\begin{equation}
  \label{eq:rhop}
   \rho'_i(t,\textbf{v})-\frac{1}{3}\text{Tr}\rho'_i\mathbbm{1}=:\frac{g_i}{2}
\begin{pmatrix} 
\mathfrak{s}_{1}^i(t,\textbf{v}) & 
\mathcal{S}_{12}^i(t,\textbf{v}) & 
\mathcal{S}_{13}^i(t,\textbf{v})\\
 \mathcal{S}_{12}^{\ast i}(t,\textbf{v}) &  
\mathfrak{s}_{2}^i(t,\textbf{v}) & 
\mathcal{S}_{23}^i(t,\textbf{v}) \\
 \mathcal{S}_{13}^{\ast i}(t,\textbf{v}) & 
\mathcal{S}_{23}^{\ast i}(t,\textbf{v}) & 
- \mathfrak{s}_{1}^i(t,\textbf{v})- \mathfrak{s}_{2}^i(t,\textbf{v})
\end{pmatrix}
\end{equation}
Here, we already used that the diagonal is unchanged by the transformation to first order in $\omega/\lambda$. The off-diagonal elements transforms to first order as
\begin{equation}
  \begin{aligned}
    \mathcal{S}_{12}^i(t,\textbf{v}) &\approx \mathcal{S}_{ex}^i(t,\textbf{v}) - \frac{\omega}{\lambda} s_{12} c_{12} c_{13} (\mathfrak{s}_1^i(t,\textbf{v}) - \mathfrak{s}_2^i(t,\textbf{v}) ), \\
    \mathcal{S}_{13}^i(t,\textbf{v}) &\approx \mathcal{S}_{ey}^i(t,\textbf{v}) + \frac{\omega}{\lambda} ( \eta s_{13} c_{13} + s_{12}^2 s_{13} c_{13} ) (2 \mathfrak{s}_1^i(t,\textbf{v}) + \mathfrak{s}_2^i(t,\textbf{v}) ), \\
    \mathcal{S}_{23}^i(t,\textbf{v}) &\approx \mathcal{S}_{xy}^i(t,\textbf{v}) .
  \end{aligned}
\end{equation}
For the antineutrinos, the transformation is similar, but with opposite signs for the terms proportional to $\omega/\lambda$. 

Using \eref{eq:Hdiag} to derive the Vacuum-Term (I) and the Matter-Term (II) from \eref{eq:Goal}, we recover the Matter-Term in \eref{eq:MatterTerm}. For the Vacuum-Term in \eref{eq:VacTermDec}, $\underline{\underline{A_0}}$ and $\underline{\underline{B_0}}$ become diagonal, while the remaining four matrices $\underline{\underline{A_1}}$, $\underline{\underline{B_1}}$, $\underline{\underline{A_2}}$ and $\underline{\underline{B_2}}$ vanish.
The result of this simplification is that $\mathcal{S}_{12}^i$, $\mathcal{S}_{13}^i$ and $\mathcal{S}_{23}^i$ decouple, and we can consider them one by one -- effectively in two-neutrino setups. 

The vacuum term now gives
\begin{equation}
  \label{eq:A0B0}
  \omega_i (\underline{\underline{A_0}} + \eta \underline{\underline{B_0}}) = \omega_i \; \textrm{diag}\left( - c_{12}^2 + c_{13}^2s_{12}^2 + \eta s_{13}^2, C_{13} (s_{12}^2 - \eta ), c_{12}^2 - s_{12}^2s_{13}^2 - \eta c_{13}^2 \right) .
\end{equation}
If the two-neutrino limit is realised by putting $\theta_{12} = \theta_{13} = 0$, the diagonal matrix has the entries $-\omega_i$, $- \omega_i \eta$ and $-\omega_i (\eta-1)$. This corresponds to three pairs of two-neutrino oscillations with normal ordering if $\eta > 0$ and inverted ordering for the last two if $\eta < 0$. However, the measured values of $\theta_{12}$ and $\theta_{13}$ are non-zero, and plugging in the values reveals that $- c_{12}^2 + c_{13}^2s_{12}^2 + \eta s_{13}^2 > 0$ for $\eta > 0$. Hence \emph{normal mass ordering gives rise to an effective two-neutrino system with inverted mass ordering}.
This is exactly what we have seen in \fref{fig:3NNO} and \fref{fig:3NNOTest}: Even in the case of normal mass ordering, an instability is present that belongs to the symmetric modes which is typical for a two-neutrino system with inverted mass ordering.

With the understanding that the three-neutrino instability we have identified can be interpreted in the two-neutrino approximation by using 
\begin{equation}
\label{eq:omegaeff}
\omega_{\textrm{eff}} =  \omega( - c_{12}^2 + c_{13}^2s_{12}^2 + \eta s_{13}^2)  ,
\end{equation}
we can expand our results by considering previous results on linear stability in the literature. Specifically, in the bulb model, the results from \cite{Chakraborty:2015tfa} on multi-angle effects and small scale inhomogeneities can be directly applied to the three-neutrino instability by using $\omega_{\textrm{eff}}$.
The absolute value of $\omega_{\textrm{eff}}$ is significantly smaller than $\omega\eta$, and the growth rate is thereby also reduced. Although this will delay the flavour conversion, it does not affect the stability as such. When comparing the results for NO and IO in \cite{Chakraborty:2015tfa}, there are regions of the $(\mu,\lambda)$ plane where IO is unstable and NO is stable. Consequently, for a supernova density profile transversing these regions, we expect to see collective oscillations that would only be predicted by a simple two-neutrino analysis if $\omega_{\textrm{eff}}$ from \eref{eq:omegaeff} was properly used.

\section{Conclusion}
\label{sec:conclusion}
In this work, we studied self-induced flavor conversion in SNe taking into account all three neutrino flavors. We constructed an approximate solution to the non-linear Von Neumann equation which describes the time dependent propagation of flavor density matrix and includes mixing among different neutrino modes. The ensemble of neutrinos were assumed to be homogeneous and collisonless, and we neglected the CP-violating phase in the mixing matrix. In the vacuum term of the the system Hamiltonian, we dropped terms that remain constant to linear order in $\mathcal{S}_{\alpha\beta}$ and that are suppressed in the environment of a large matter potential. The commutator of different neutrino modes was calculated to first order in $|\mathcal{S}_{\alpha\beta}|$, and only a discrete number of different modes was assumed. Finally, the differential equation was transformed into an eigenvalue problem by making a plane wave ansatz for $\mathcal{S}_{\alpha\beta}$.

We applied our approximate description to a system of two colliding beams of neutrinos and antineutrinos and found flavor instabilities not only for IO but also for NO. In order to get a deeper understanding, we turned to a model of four colliding neutrino/antineutrino beams that intersect with an angle $\alpha$. In this system, we studied symmetric and antisymmetric modes and found that symmetric modes can develop flavor instabilities for both mass orderings. The origin of these instabilities in NO is the value of the mass square ratio $\eta$ combined with those of the mixing angles $\theta_{12}$ and $\theta_{13}$.

Finally, we showed in the framework of propagating states that the new instability for NO can be interpreted as an effective two-neutrino system with $\omega_{\textrm{eff}}<0$ corresponding to an effective IO. With this understanding, our results are expected to be valid beyond our simple toy systems as one simply replaces $\omega$ in any two-neutrino analysis by $\omega_{\textrm{eff}}$.

With a good understanding of all the instabilities that we identify, we can also confirm that the off-diagonal terms of the density matrix do not lead to any new instabilities at least when the CP-violating phase is neglected and equal fluxes of muon and tau neutrinos are assumed.

Our findings show that instabilities that were only expected to exist for IO are also present in NO, albeit with a smaller growth rate. This once again demonstrates how great care must be taken when dealing with collective neutrino oscillations in supernova in order to capture all the relevant physics.

\section{Acknowledgements}
The authors thank Alexei Yu Smirnov for fruitful discussion on the topic of self-induced neutrino conversion. R.S.L.H. was partly funded by the Alexander von Humboldt Foundation.

\bibliographystyle{JHEP}
\bibliography{Literatur}
\end{document}